\documentclass[preprint,pra,aps,preprintnumbers,amsmath,amssymb,eqsecnum]{revtex4}

\usepackage{graphicx}
\usepackage{dcolumn}
\usepackage{bm}
\usepackage{verbatim}
\usepackage{epsfig}
\usepackage{epstopdf}

\newcommand\beq{\begin{equation}}
\newcommand\eeq{\end{equation}}
\newcommand\bea{\begin{eqnarray}}
\newcommand\eea{\end{eqnarray}}

\def\half{\frac {1} {2}}

\def\x0{{{\bf x}_0}}

\begin{document}


\title{Leggett-Garg tests of macrorealism: checks for non-invasiveness and generalizations to 
higher-order correlators}


\author{J.J.Halliwell}%
\email{j.halliwell@imperial.ac.uk}

\affiliation{Blackett Laboratory \\ Imperial College \\ London SW7 2BZ \\ UK }



\begin{abstract}
In the tests for macrorealism proposed by Leggett and Garg, the temporal correlation functions of a dichotomic variable $Q$ must be measured in a non-invasive way to rule out alternative classical explanations of Leggett-Garg inequality violations.
Ideal negative measurements, in which a null result is argued to be a non-invasive determination of the system's state, are often used.
From a quantum-mechanical perspective, such a measurement collapses the wave function and will therefore typically be found to be invasive under any experimental check.
Here, a simple modified ideal negative measurement protocol is described for measuring the correlation functions which is argued to be non-invasive from both classical and quantum perspectives and hence the non-invasiveness
can then be checked experimentally, thereby permitting a quantitative measure of the degree of clumsiness of the measurement. It is also shown how this procedure may be extended to measure higher-order correlation functions and a number of higher-order conditions for macrorealism are derived.
\end{abstract}

\maketitle

\section{Introduction}

A number of experiments in recent years have aimed to test the world view known as ``macroscopic realism'' (macrorealism), the view that a system evolving in time can be regarded as possessing definite properties at each time, regardless of past or future measurements. This view, first put forwards by Leggett and Garg \cite{LG1,L1}, is made precise by breaking macrorealism (MR) into the following three assumptions:  
the system is in one of the states available to it at each moment of time  (macrorealism {\it per se}, MRps); it is possible in principle to determine the state of the system without disturbing the subsequent dynamics (non-invasive measurability, NIM);
future measurements cannot affect the present state (induction).

MR is traditionally investigated
using a single dichomotic variable $Q$ which is measured in a number of different experiments involving three (or more)  pairs of times thereby determining a set of temporal correlation function of the form,
\beq
C_{12} = \langle Q(t_1) Q(t_2) \rangle,
\label{corr}
\eeq
and also the averages of $Q$ at each time, $\langle Q_1 \rangle$ etc, where $Q_1$ denotes $Q(t_1)$. For a macrorealistic  theory, the above three assumptions imply the existence of a joint probability distribution for $Q$ at the three times $t_1, t_2, t_3$ and it readily follows from this that the temporal correlation functions obey the Leggett-Garg (LG) inequalities:
\bea
1 + C_{12} + C_{23} + C_{13} & \ge & 0,
\label{LG1}
\\
1 - C_{12} - C_{23} + C_{13} & \ge & 0,
\label{LG2}
\\
1 + C_{12} - C_{23} - C_{13} & \ge & 0,
\\
1 - C_{12} + C_{23} - C_{13} & \ge & 0.
\label{LG4}
\eea
These inequalities are necessary conditions for macrorealism, but not sufficient. They may however be turned into a sufficient set of conditions by adjoining them with a set of twelve two-time LG inequalities, of the form
\bea
1 +  \langle Q_1 \rangle + \langle Q_2 \rangle + C_{12} &\ge& 0, \\
1 -  \langle Q_1 \rangle - \langle Q_2 \rangle + C_{12} &\ge& 0, \\
1 +  \langle Q_1 \rangle - \langle Q_2 \rangle - C_{12} &\ge& 0, \\
1 -  \langle Q_1 \rangle + \langle Q_2 \rangle - C_{12} &\ge& 0,
\label{LG2}
\eea
plus two more sets of four inequalities, at the time pairs $(t_2,t_3)$ and $(t_1, t_3)$ \cite{HalQ,HalLG4}. These sixteen inequalities are a sufficient set since they have a  mathematical parallel with the Bell case and Fine's theorem then applies \cite{Fine,HalFine}. A decisive test of macrorealism for measurements at three times entails measurement of the three correlation functions and three averages (in six different experiments, one for each) and checking the sixteen LG inequalities. Almost all tests of the LG inequalities check only a subset of the above sixteen inequalities, although there is a recent attempt to test all sixteen \cite{Laf}.
Many experimental tests and theoretical aspects of the LG inequalities  
are reviewed in Ref.\cite{ELN} (and see also Ref.\cite{MaTi} for a critique and clarification of what the LG inequalities actually test).

Recently, an alternative approach to characterizing macrorealism has been proposed which consists of determining the underlying probability using a single experiment in which $Q$ is measured sequentially at all three times \cite{KoBr,Cle}. (See also Refs.\cite{MaTi,Ma1}).
We denote this probability by $p_{123} (s_1, s_2, s_3) $, which is the probability that $Q_1$ takes the value $s_1 = \pm 1 $ etc. This probability is then compared with probabilities obtained the same way in experiments involves measurements at one and two times and a series of ``no-signaling in time'' (NSIT) conditions are imposed, which for one and two-time measurements have the form,
\beq
\sum_{s_1} p_{12} (s_1,s_2) = p_2( s_2).
\label{NSIT12} \\
\eeq
These, together with further similar conditions on the three-time probability 
ensure that such sequential measurements are non-invasive. These conditions are much stronger than the LG inequalities (and are related to ``coherence witness" conditions \cite{Wit,ScEm}).
As argued in Ref.\cite{HalLG4}, they entail a stronger notion of NIM and hence of macrorealism. 
Mention should also be made of the Wigner-Leggett-Garg inequalities which lie midway between traditional LG tests and NSIT conditions \cite{WLG}.

The first aim of the present paper is to focus on how to satisfy the NIM condition, which is the most tendentious issue around tests of the LG inequalities. 
Leggett and Garg proposed that the measurement of the correlation functions be carried out using ideal negative measurements, in which the detector is coupled to, say, only the $Q=+1$ state, at the first time, and a null result then implies that the system is in the $Q=-1$ state but without any interaction taking place. The experiment is then repeated with the coupling to the $Q=-1$ state.
This procedure rules out alternative classical explanations \cite{Mon,Ye1,Guh} and has been successfully implemented in a number of recent experiments \cite {Knee,Rob,KBLL,EmINRM}. 
Alternative protocols for implementing (or modifying) NIM  have also been proposed \cite{HalLG2,HalLG3,Ema,Hue,FA,AF,Lap,Zuk}. Many other experimental tests of the LG inequalities have also been carried out, on a variety of different physical systems (see for example the extensive list of references in Refs.\cite{ELN,HalLG4}).

Although thorough arguments for non-invasiveness using ideal negative measurements are given in the reports of experimental tests of the LG inequalities, what is clearly desirable is an {\it experimental check} of invasiveness. A purely classical system measured using an ideal negative measurement could determine a two-time probability $p_{12}(s_1,s_2)$ and we would expect it to obey the NSIT condition, Eq.(\ref{NSIT12}). It would only fail to satisfy this condition if there was some inadvertent ``clumsiness'' in the experimental procedure. Such clumsiness is always present in any realistic experimental arrangement but conditions such as Eq.(\ref{NSIT12}) can be employed to make sure that clumsiness is kept sufficiently small \cite{deco1,deco2}.

However, experimental devices obey the laws of quantum mechanics and the above classical argument does not necessarily apply. The problem is that ideal negative measurements still collapse the wave function \cite{Dicke} and as result, the NSIT condition Eq.(\ref{NSIT12}), which is sensitive to interference, will not be satisfied in general, even for the most skillfully conducted ideal negative measurement. This means that  the NSIT condition Eq.(\ref{NSIT12}) is not necessarily a good detector of clumsiness since it cannot distinguish clumsiness from collapse of the wave function effects.
These observations by no means undermine any of the tests of the LG inequalities that use ideal negative measurements, but any observed violation of the LG inequalities will clearly be a more 
convincing refutation of macrorealism if it is accompanied by an experimental confirmation of no signaling.
It is therefore useful to find a modified ideal negative measurement protocol which separates collapse of the wave function effects from experimental clumsiness and enables NSIT conditions to be put to use as practically useful checks of NIM .

Simple experimental checks of NIM were first suggested by Leggett  in Ref.\cite{L1}. Wilde and Mizel defined an ``adroit measurement'', using conditions which, in the language of the present paper, are approximate NSIT conditions \cite{deco1}. (This approach was tested in Ref.\cite{deco2}).
Katiyar et al \cite{KBLL} and Knee et al \cite{Knee2} used control experiments to address clumsiness loophole (which consisted of assessing the degree of invasivess for a set of control states in which $Q$ takes a definite value).
George et al \cite{3box} and Emary \cite{Ema} exhibited situations in which an LG inequality violation is accompanied by certain NSIT conditions being satisfied, in a three-level system.

In the present paper, it will be shown how the usual ideal negative measurement protocol may be modified in a simple way so that a NSIT condition should be satisfied for a properly executed experiment for arbitrary initial states and arbitrary choices of times of measurements. 
It will not be necessary to choose special initial states of choices of times to satisfy NSIT
(in contrast, for example, to Refs.\cite{deco1,3box,Ema}). 
We will focus on the definition of macrorealism described above which uses an extended set of LG inequalities, with NSIT conditions used purely as checks on NIM (and not as definitions of macrorealism itself).


The second aim of this paper is to consider how the LG framework is generalized when higher order correlation functions are involved. This has become interesting of late since experiments have been done which can measure third and higher order correlators (see, for example, Ref.\cite{Bec}).
In particular we will write down necessary and sufficient conditions for macrorealism which include third and fourth order correlators. This development is logically separate from the first aim of this paper. However, it clearly provides a natural challenge for extending the modified ideal negative measurement protocol to a more complicated situation, and we show how to do this.

The modified ideal negative measurement protocol is described in Section 2. In Section 3, LG tests involving higher order correlators are considered. Measurement of such correlators using the modified ideal negative measurement protocol is described in Section 4. We summarize in Section 5.
Some useful quantum-mechanical results are outlined in Appendix A.

\section{A Modified Ideal Negative Measurement Protocol}


\subsection{The Approach}

We consider a standard LG test involving the two and three-time LG inequalities described above.
We will assume that the system starts out at $t=0$ in an initial state described by a density operator $\rho$ (which may be unknown) and measurements of $Q$ are made in a set of experiments at one or two times chosen from the set $ t_1, t_2, t_3 $ (where $0<t_1<t_2<t_3$).
By doing three experiments with a single-time measurement in each, we may determine the three averages,  $\langle Q_1 \rangle$,  $\langle Q_2 \rangle$ and  $\langle Q_3 \rangle$. There is no issue of invasiveness since only a single measurement is made in each experiment.

The three correlation functions are determined in another three experiments with a pair of measurements made in each, using ideal negative measurements. The question we address is how to modify the ideal negative measurement protocol so that it is non-invasive as defined by a NSIT condition.
Our approach is to consider modifications which consist of
some operation carried out just before the time of the first measurement in each pair which would be acceptable to a macrorealist and which could potentially improve the situation with regard to invasiveness.

\subsection{Quantum-Mechanical Analysis}

As discussed the reason for the failure of NSIT in general is quantum-mechanical in nature so a quantum-mechanical analysis is required. However, it must of course ultimately be phrase in macrorealistic terms and we will do this at the end of this section.


For a system in initial state $\rho$ at $t=0$ the quantum-mechanical probability for a single time measurement at time $t_1$  is
\beq
p_1 (s_1) = {\rm Tr} \left( P_{s_1} (t_1) \rho \right)
\label{single}
\eeq
where the projection operators $P_s (t) $ are defined by $P_s (t) = e^{i Ht} P_s  e^{ -i H t } $ (in units in which $\hbar=1$) and
\beq
P_s = \half \left( 1 + s \hat Q \right).
\eeq
The probability for sequential projective measurements at times $t_1, t_2$ is,
\beq
p_{12}(s_1,s_2) = {\rm Tr} \left( P_{s_2} (t_2) P_{s_1} (t_1) \rho P_{s_1} (t_1) \right).
\label{2time}
\eeq
Note that this expression involves only the terms $P_+(t_1) \rho P_+ (t_1)$ and $P_-(t_1) \rho P_- (t_1)$ in $p_{12}(s_1,s_2)$, which means that it depends only on the diagonal part of 
$\rho(t_1) = e^{-iHt_1} \rho  e^{iHt_1}$, which we denote $\rho_{\rm diag} (t_1)$ and may be written,
\beq
\rho_{\rm diag} (t_1)  = \sum_{s_1} P_{s_1} \rho(t_1)  P_{s_1}.
\eeq
(That is, $\rho_{diag}$ is an initial density matrix, not necessarily diagonal, which evolves to become equal to the diagonal part of $\rho$ in the $Q$ basis at $t_1$).
This also means that the
temporal correlation function 
\beq
C_{12} = \sum_{s_1, s_2} s_1 s_2 \ p_{12}(s_1, s_2 ),
\label{sameC}
\eeq
is unchanged by replacing  $\rho(t_1) $ with $\rho_{\rm diag} (t_1) $. (In fact in the simplest two-state systems the correlation function is completely independent of the initial state).

Summing over the initial measurement we find
\beq
\sum_{s_1} p_{12} (s_1, s_2 ) = {\rm Tr} \left( P_{s_2} (t_2)  \rho_{\rm diag}  \right).
\label{con}
\eeq
The right hand side of Eq.(\ref{con}) is not in general equal to the single time measurement result, 
\beq
p_2 (s_2) = {\rm Tr} (P_{s_2} (t_2) \rho ),
\eeq
hence the NSIT condition Eq.(\ref{NSIT12}) is not satisfied. This is readily seen to be related to interference between different histories of the system, as outlined in Appendix A.

However, NSIT clearly {\it will} be satisfied if $\rho(t_1) = \rho_{\rm diag} (t_1)$.
This suggests the following strategy for determining the correlator $C_{12}$ (and only the correlator) using a method that satisfies NSIT.
First, ideal negative measurements are used to determine $p_{12}(s_1,s_2)$, with an initial state $\rho$, and the correlation function $C_{12}$ is read off. The probability for $Q$ at time $t_2$ only $ p_2(s_2)$ is determined in a seperate, single measurement experiment and the NSIT condition is checked. It will generally be found to fail (but if not, no further steps are required).

We then do a different experiment in which an operation is carried out immediately before the first measurement, consisting of
a rapidly acting diagonalization procedure (about which more below) shortly before time $t_1$, which has the effect of replacing $\rho(t_1) $ with its diagonal counterpart $\rho_{\rm diag} (t_1)$. The two-time probability $p_{12}(s_1,s_2)$ is again measured with this different initial state, and checked to see that it gives the same result as the original experiment, as predicted by quantum mechanics.
 In particular the correlators should be the same. 
(This check will confirm that the diagonalizing mechanism is not doing anything spurious). Again a 
single-time measurement at $t_2$ in a separate experiment is carried out, but now with the diagonalizing mechanism just before $t_1$ still in place,
yielding a result 
\beq
\tilde p_2(s_2) = {\rm Tr} \left( P_{s_2} (t_2) \rho_{\rm diag} \right).
\eeq
Crucially, this now means that the NSIT condition Eq.(\ref{con}) is satisfied.
(Note that $ \tilde p_2(s_2) $ is of course different to $p_2(s_2)$ but this does not matter since we are not using this experiment to determine $\langle Q_2 \rangle$).

In brief, because the correlation function is insensitive to diagonalization at time $t_1$,
we can replace the original situation of interest that does not satisfy NSIT with a very similar situation with the same correlation function but which
does satisfy NSIT. 
In essence, we have classicalized the system in terms of its behaviour at two times, which means that, for the purposes of measuring the correlator only, the classical and quantum descriptions of ideal negative measurements coincide.  
Since NSIT is in principle satisfied, for a perfectly executed experiment, this means that any detected violation of it must come not from wave function collapse but from experimental clumsiness.
This is therefore the sought-after check on the non-invasiveness of ideal negative measurements of the correlation function.



\subsection{The Diagonalization Procedure}

The diagonalization procedure can be carried out in a number of different ways. Of course it is generally known that coupling a system to some sort of environment which is averaged out tends to create a situation in which the density matrix tends towards diagonality.
One could simply couple the system to an environment, for the entire time period of interest, but this would also change the correlation functions and significantly lessen the LG inequality violation. A brief but very efficient period of decoherence just before the time of the first measurement in each pair is what is required.

Two such methods which have been used experimentally are conveniently summarized in Ref.\cite{Knee3}.  One is ``artificial dephasing'' in which a random distribution of phase factors are applied to the density operator and then averaged over \cite{Wang}.
A second possibility is a ``blind measurement" \cite{ScEm}, in which the density operator is measured in the preferred basis (i.e eigenstates of $Q$) at the first time and then the result is simply discarded.

As it happens, some of the recent experimental tests of the LG inequalities which implement ideal negative measurements, use an ancilla system  \cite {Knee,Rob,KBLL,PaMa} which could easily be utilized to carry out a  blind measurement. In these approaches, the primary system at time $t_1$ is interacted with an ancilla using a controlled NOT gate, and the state of the total system immediately afterwards has the form,
\beq
| \Psi_T \rangle =  P_+ e^{-iHt_1} |\psi \rangle \otimes |a_+ \rangle + P_- e^{-iHt_1} | \psi \rangle \otimes |a_- \rangle,
\eeq
where we have taken a pure initial state $|\psi \rangle$ for the primary system and $|a_{\pm} \rangle$ denote the ancilla states. A measurement of the primary system is subquently made at time $t_2$ along with a measurement of the ancilla (whose state remains constant for $t>t_1$). Since the ancilla states $ |a_{\pm} \rangle $ are perfectly correlated with $s_1$, these measurements thus determine $p_{12}(s_1,s_2)$.

To check NSIT, a measurement is made of the primary system only at time $t_2$ with the ancilla still coupled in but its result discarded.
This means that the system is then described by the mixed state obtained by tracing out the ancilla:
\beq
\rho_S =   P_+ |\psi_{t_1} \rangle \langle \psi_{t_1} |  P_+ +  P_- |\psi_{t_1} \rangle \langle \psi_{t_1} |  P_-
\eeq
where $| \psi_{t_1} \rangle = e^{-iHt} | \psi \rangle$.
This is, as desired, diagonal in $Q$ at time $t_1$ which means that the NSIT condition 
will be satisfied. 
This method of implementing the modified ideal negative measurement protocol is perhaps the most practically useful one since it makes use of existing techniques.

\subsection{Macrorealistic Formulation}

The above analysis is quantum-mechanical in nature and
it is important in the LG framework to phrase things in macrorealistic terms.
A convenient way to do this is to simply rephrase the above analysis as a testable assumption, about an ideal negative measurement of the correlation function, namely:
{\it there exists an operation which acts very briefly just before the first measurement time which does not change the value of the correlation function, and, for which the resultant two-time probability is, in principle, compatible with the NSIT condition}. Here, ``in principle" means for a perfectly executed experiment. This assumption clearly provides a justification for using the NSIT condition as detector of clumsiness.


The strength of this assumption is that it is actually true from the quantum-mechanical perspective.
Furthermore, the macrorealist , who in effect sees only diagonal density matrices so is indifferent to the diagonalization process, would find the first part of the assumption very plausible. 
The macrorealist would also expect NSIT to hold since the measurement is an ideal negative measurement.  Hence both parts of the assumption are macrorealistically reasonable. Moreover they can both be checked experimentally.

A possible objection to the above procedure concerns the degree to which NSIT conditions really characterize non-invasiveness.
They were originally described as a statistical version of NIM \cite{KoBr} and with a mixed initial state can be satisfied essentially by averaging two situations which are individually invasive \cite{SOS,KGBB}. However, this is not a concern here since we are using ideal negative measurements which are ontically non-invasive.

\subsection{Generalizations to Many-Valued Measurements}

The approach above is given for a measurements of a single dichotomic variable $Q$ but some LG tests entail measurements onto three or more alternatives at each time, labeled $n$ say, where $n =1, 2 \cdots N$ to yield a two-time probability $p_{12} (n_1, n_2)$. For examples Refs.\cite{3box,Ema} considered a three-state system. The two-time probability still has the general form Eq.(\ref{2time}) in quantum mechanics which means that the correlation functions are again insensitive to diagonalization of the density matrix at the first time. However, the situation with regard to NSIT conditions is more complicated. The natural NSIT to consider is
\beq
\sum_{n_1} p_{12} (n_1, n_2) = p_2 (n_2),
\label{NSITn}
\eeq
and this consists of two independent conditions. However, there are other types of NSIT condition obtained, for example, by constructing a single dichotomic variable  $Q$ at the first time, with values $s_1= \pm1 $, and measuring only this (rather than all values of $n_1$), to yield the probability $\tilde p_{12} (s_1, n_2)$. The corresponding NSIT condition is
\beq
\sum_{s_1} \tilde p_{12} (s_1, n_2) = p_2 (n_2).
\eeq
In a macrorealistic theory these different types of NSIT conditions are trivially related but not so in quantum mechanics. This leads to some interesting new features, for example, correlation functions which can violate the Tsirelson inequalities \cite{EmINRM}, and to situations in which there are violations of the two-time LG inequalities but the NSIT condition Eq.(\ref{NSITn}) is still satisfied \cite{3box,Ema}. 

A more thorough discussion of NSIT and other conditions in this situation will be given elsewhere.  What is clear from the form of the above NSIT conditions is that the basic method of causing NSIT conditions to be satisfied using a diagonalization procedure will still hold. Hence the modified ideal negative measurement protocol for measuring correlation functions may be applied here.


\section{Leggett-Garg Tests of Macrorealism with Higher Order Correlation Functions}

Another natural generalization to consider is measurement of higher order correlation functions, but first it is useful to develop conditions for macrorealism that involve them.
Although the Leggett-Garg framework was originally introduced in the context of experiments in which measurements are made of $Q$ at pairs of times chosen from three or four possible times, the most general possible situation in which macrorealism could be tested, for a dichotomic variable, involves a set of $n$ possible times and a set of measurements made in each experiment at $m$ times, where $m \le n$. The case $m=2$ for arbitary $n$ involves $n$-time LG inequalities, considered, for example, in Refs.\cite{LGn1,LGn2} (reviewed in Ref.\cite{ELN}).
The case $ m > 2 $ does not appear to have been considered in any detail in a Leggett-Garg framework,
although measurements of higher moments have been discusssed \cite{HOM} and a recent LG-type experiment to test them has been conducted \cite{Bec}.

Of course, in the stronger tests for macrorealism defined purely in terms of NSIT in conditions outlined above \cite{KoBr,Cle},
one could quite simply use sequential measurements at $n$ times and then impose a series of NSIT conditions on the resulting measured probability $p_{12\cdots n} (s_1, s_2, \cdots s_n)$. Here, however, we are interested in the weaker notion of macrorealism which entails measuring correlation functions in a series of different experiments and then seeking LG-type conditions under which they can be assembled into a unifying probability. 


\subsection{Measurements at Three Times}

We consider first the case of measurements involving just three times. This turns out to be quite simple. We suppose that measurements have been used to determine
the averages at one time $\langle Q_i \rangle $ (where $i=1,2,3$), the correlators at two times $C_{ij}$ (where $ij =12,23,13$), but also a single third-order correlation function,
\beq
D_{123} = \langle Q_1 Q_2 Q_3 \rangle.
\eeq
(where it is assumed that all seven moments are measured in seven different experiments).
This case is simple because the above moments together fix the three-time candidate probability uniquely to be,
\bea
p(s_1,s_2,s_3) &=& \frac{1}{8} \left( 1 + s_1 \langle Q_1 \rangle + s_2 \langle Q_2 \rangle +s_3 \langle Q_3 \rangle 
\right.
\nonumber \\
&+&  \left. s_1 s_2 C_{12} + s_2 s_3 C_{23} + s_1 s_3 C_{13} + s_1 s_2 s_3 {D_{123}}
\right) .
\label{p123}
\eea
 It is a candidate probability since an expression of the form 
Eq.(\ref{p123}) constructed from a set of moments
is not necessarily non-negative, and indeed we expect that it could be negative under certain circumstances in quantum mechanics.


However, for a macrorealistic theory, a joint probability on the variables $Q_1$, $Q_2$, $Q_3$ must exist and $p(s_1,s_2,s_3)$ may then be written,
\beq
p(s_1,s_2,s_3) = \frac{1}{8} \langle ( 1 + s_1 Q_1 ) (1 + s_2 Q_2) ( 1+ s_3 Q_3) \rangle,
\eeq
which is manifestly non-negative.
This means that the set of eight conditions
\beq
p(s_1,s_2,s_3) \ge 0,
\label{3tp}
\eeq
form a necessary and sufficient set of conditions for macrorealism.
The contrast with the usual LG framework involving two and three-time inequalities is that the correlator $D_{123}$ is not fixed and the sixteen (two and three-time) LG inequalities are the necessary and sufficient condition that there exists {\it some} value of $D_{123}$ for which $p(s_1,s_2,s_3) \ge 0$.


\subsection{Measurements at More than Three Times}

For measurements at $n$ times, if all possible correlation functions are determined the candidate probability is again uniquely determined and
the above condition readily generalizes to,
\beq
p(s_1, s_2 \cdots s_n) = \frac{1} {2^n} \langle \prod_{i=1}^n  (1 + s_i Q_i) \rangle \ge 0.
\label{ntimes}
\eeq
As argued in Appendix A, quantities of this general type can be negative in quantum mechanics.

The $n=4$ case is given explicitly by,
\beq
p (s_1, s_2, s_3, s_4) = \frac {1} {16} \left( 1 +
\sum_i s_i \langle Q_i \rangle  + \sum_{ij} s_i s_j C_{ij}  + \sum_{ijk} s_i s_j s_k D_{ijk},
+ s_1 s_2 s_3 s_4 E \right),
\label{qs}
\eeq
where the indices $i,j,k$ run over the values $1,2,3,4$, the summation over $ij$ has $i < j$ and the summation over $ijk$ has $i<j<k$. There are six second-order correlation functions $C_{ij}$, four third-order correlation functions,
\beq
D_{ijk} = \langle Q_i Q_j Q_k \rangle,
\eeq
and one fourth-order correlation function
\beq
E = \langle Q_1 Q_2 Q_3 Q_4 \rangle.
\eeq
The usual LG scenario in this case involves measurement of the four averages $\langle Q_i \rangle$ and four of the six correlation functions, $C_{12}, C_{23}, C_{34}, C_{14}$.
The necessary and sufficient conditions for MR are then the sixteen two-time LG inequalities and the eight four-time LG inequalities \cite{HalLG4}.

One can imagine more complicated situations in which some, but not all, of the higher-order correlation functions are measured. The general procedure employed here is a generalization of a simple method described in Ref.\cite{HalFine}. Requiring that Eq.(\ref{qs}) is non-negative for all values of $s_1, s_2, s_3, s_4$ yields a set of lower and upper bounds on the unfixed correlators, where the lower and upper bounds depend on the fixed quantities. The unfixed quantities may then be chosen so that Eq.(\ref{qs}) is non-negative as long as all of their upper bounds are greater or equal to all of their lower bounds. This yields a set of conditions on the fixed quantities which ensure that the unifying probability may be constructed.

Note however, that in the case in which not all correlators are fixed,
there can exist simplifications in which the existence or not of a four-time probability reduces to a set of conditions on some three-time probabilities. Suppose for example, that measurements have been made that completely fix the two three-time probabilities $p(s_1,s_2,s_3)$ and $p(s_1,s_2,s_4)$ and, they are non-negative. This fixes all the averages and all correlators except $C_{34}$, $D_{234}$, $D_{134}$ and $E$. One could follow the above general procedure to determine whether values of the unfixed correlators can be chosen so that there exists a non-negative four-time probability, $p(s_1,s_2,s_3,s_4)$ which matches the fixed three-times ones.
However, it turns out that there is a much simpler way, which is to use the ansatz first introduced by Fine \cite{Fine}, and note that the solution is,
\beq
p(s_1,s_2,s_3,s_4) = \frac{ p(s_1,s_2,s_3) p(s_1,s_2,s_4) } { p_{12}(s_1,s_2) },
\eeq
since summing out $s_3$ or $s_4$ is readily seen to yield $p(s_1,s_2,s_4)$ and $p(s_1,s_2,s_3)$, respectively. This suggests that the most interesting cases involving higher-order correlators are those in which all the correlators are known.



\subsection{A Different Type of MR Condition}

We also note that there is a different type of condition involving measurements at arbitrary numbers of times which is neither a NSIT condition, nor a LG test. Suppose sequential measurements are made at, say, three times, to determine $p_{123} (s_1,s_2,s_3)$ and this is compared with the two-time result, $p_{23} (s_2,s_3)$. For a macrorealistic theory we expect that a NSIT condition of the form
\beq
p_{23} (s_2, s_3) = \sum_{s_1} p_{123} (s_1, s_2,s_3),
\eeq
should hold. But since every term on the RHS is non-negative, this implies that,
\beq
p_{123} (s_1,s_2,s_3) \le p_{23} (s_2, s_3),
\eeq
for any $s_1$.
Similarly, for two-time measurements, a macrorealistic theory should satisfy,
\beq
p_{12} (s_1, s_2) \le p_2 (s_2).
\label{ineqp12}
\eeq
Generalization to arbitrary numbers of measurements is obvious.
Conditions of this type were tested experimentally in Ref.\cite{Bec}. They do not seem to have been investigated elsewhere, although they are very similar to the Wigner-Leggett-Garg inequalities \cite{WLG}, if not actually the same in some cases. They can be violated in quantum theory, as outlined in Appendix A.

\section{Measurement of Higher Order Correlation Functions}

We describe now the non-invasive measurement of a third order correlation function using the modified ideal negative measurement protocol. We assume that the averages $\langle Q_i \rangle $ and second order correlators $C_{ij}$ have already been determined in a set of separate experiments. The third order correlator is needed in order to check whether the condition Eq.(\ref{3tp}) holds.

The approach is to use two ideal negative measurements
to non-invasively determine the three-time sequential measurement probability $p_{123} (s_1, s_2, s_3)$, from which the third order correlator,
\beq
D_{123} = \sum_{s_1 s_2 s_3} s_1 s_2 s_3 \ p_{123} (s_1, s_2, s_3),
\eeq
is readily obtained. The three-time probability $p_{123} (s_1,s_2,s_3)$ is used {\it only} to determine the third order correlator and not to determine any of the lower moments, which are determined already.

Two different ideal negative measurements are required involving two detectors.
In the first measurement, the first detector is coupled to, say, the $Q=+1$ state at time $t_1$ and the second detector to the $Q=+1$ state at time $t_2$. We denote this the $(+,+)$ configuration.
A projective measurement of the system is also made at time $t_3$, with result $s_3$. We are interested only in the situation in which neither detector triggers. All other situations are discarded. The fraction of runs in which neither triggers determine the probabilities $p_{123} (+,+,s_3)$.
This procedure is then repeated three more times,  for the $(+,-)$, $(-,+)$ and $(-,-)$ configurations, thereby determining the corresponding probabilities. The combination of the four experiments yields the probability $p_{123} (s_1,s_2,s_3)$.

Actually, knowledge of the the averages $\langle Q_i \rangle $ and second order correlators $C_{ij}$
means that we only need to know just one component of the three-time probability, such as $p_{123} (+,+,+)$, to read off the third order correlation via Eq.(\ref{p123}). However, we need all components of the three-time probability in order to check the NSIT conditions.

A complete set of NSIT conditions \cite{KoBr,Cle} for the three-time probability is
\bea
\sum_{s_2} p_{23} (s_2,s_3) &=& p_3( s_3),
\label{NSIT12} \\
\sum_{s_2} p_{123} (s_1, s_2,s_3) &=& p_{13} (s_1, s_3),
\label{NSIT1n3}
\\
\sum_{s_1} p_{123} (s_1, s_2,s_3) &=& p_{23} (s_2, s_3).
\label{NSITn23}
\eea
Violations of these conditions signal invasiveness. Like the two-time situation, these conditions will not be satisfied in general under ideal negative measurements, due to the presence of interferences, so we use the modified protocol.


For convenience we use the ancilla approach so two ancillas are required, one for each time.
We will assume, in analogy to the two-time case, that $p_{123} (s_1,s_2,s_3) $ is independent of whether or not the result of the measurement performed by the ancilla is discarded.
Like the two-time case, this assumption is readily seen to be true in a quantum-mechanical description -- the usual measurement formula for three times,
\beq
p(s_1, s_2, s_3) = {\rm Tr} \left( P_{s_3} (t_3) P_{s_2} (t_2) P_{s_1} (t_1) \rho P_{s_1} (t_1) P_{s_2} (t_2) \right),
\label{3time}
\eeq
is clearly independent of whether a diagonalization procedure is acting at times $t_1$ and $t_2$ (since no off-diagonal terms are involved) so likewise the third order correlator.

The NSIT condition Eq.(\ref{NSIT12}) is identical in form to that discussed in Section 2. It will therefore be satisfied if, in the measurement of $p_3 (s_3)$, the ancilla at time $t_2$ is left in place and allowed to perform a blind measurement.
To check the NSIT condition Eq.(\ref{NSIT1n3}), $p_{13} (s_1, s_3) $ is determined in a separate experiment to that determining $p_{123} (s_1,s_2,s_3)$, but again the ancilla at time $t_2$ is allowed to perform a blind measurement during the deterimination of  $p_{13} (s_1, s_3) $.
Similar Eq.(\ref{NSITn23}) is satisfied if the ancilla at time $t_1$ is allowed to perform a blind measurement during the determination of $ p_{23} (s_2,s_3)$.
In each case the blind measurement performed by the ancilla simply diagonalizes the density operator at that time and it is this that ensures that all NSIT conditions are satisfied.

We thus find the that modified ideal negative measurement protocol readily generalizes to a three-time scenario and the set of NSIT in time conditions are, for a perfectly executed experiment, satisfied. Any observed violations of them are therefore reflections of experimental clumsiness. The generalized to four or more times is clearly straightforwards, but progressively more complicated.


\section{Conclusion}

This paper addressed two aspects of Leggett-Garg tests of macrorealism. The first consisted of a modified ideal negative measurement protocol which can check for invasiveness unhindered by the effects of wave function collapse. It is based on two simple observations: firstly, that the correlation function is unchanged by digaonalizing the density operator at the first time, and secondly, that this diagonalization procedure is readily accomplished with commonly-used ancilla-based measurements in which the ancilla is used to peform a blind measurement.  A quantum-mechanical understanding of the method was presented but a macrorealistic formulation of the protocol was then given in terms of a plausible and testable assumption.

The second aspect concerned tests for macrorealism when higher order correlation functions are involved. At $n$ times, when all possible correlation functions are measured, the necessary and sufficient conditions for macrorealism consist of the set of inequalities Eq.(\ref{ntimes}). More complicated conditions arise in the event that some, but not all the higher order correlators are measured. A different type of condition for macrorealism was also described, involving comparing the probabilities for strings of sequential measurements.
It was also shown how to extend the modified ideal negative measurement protocol to measurement of higher order correlators.

Experimental implementation of the ideas presented in this paper would clearly be of interest.
This should not be too difficult by comparatively straightforward modification of existing experimental approaches.

\section{Acknowledgements}

I am very grateful Clive Emary, George Knee, Johannes Kofler, Owen Maroney and James Yearsley for many useful discussions and email exchanges about the Leggett-Garg inequalities.

\appendix

\section{Some Quantum-Mechanical Results}

As stated in Section 2, the failure of the NSIT condition is related to quantum interferences.
To see that, note that Eq.(\ref{2time}) may be written
\beq
p_{12}(s_1,s_2) = q(s_1,s_2) - \frac{1}{2} \sum_{s_1 \ne s_1'}  {\rm Re} D (s_1, s_2 |s_1', s_2)
\label{A1}
\eeq
where
\beq
D(s_1, s_2 | s_1',s_2) = {\rm Tr} \left( P_{s_2} (t_2) P_{s_1} (t_1) \rho P_{s_1'} (t_1) \right)
\eeq
is the decoherence functional, whose off-diagonal components are measures of interference between pairs of histories. Also,
\beq
q(s_1, s_2 ) = {\rm Re} {\rm Tr} \left( P_{s_2} (t_2) P_{s_1} (t_1) \rho \right),
\eeq
is a quasi-probability which formally satisfies the NSIT condition exactly, but can be negative.
(In fact, the two-time LG inequalities may be written, in quantum theory, as the inequalities $q(s_1,s_2) \ge 0 $). Hence NSIT for $p_{12}(s_1,s_2)$ fails when the off-diagonal terms of the decoherence functional are non-zero. Furthermore, it can be seen from Eq.(\ref{A1}) that, since $p_{12}(s_1,s_2)$ is always non-negative, $q(s_1,s_2)$ can be negative if the interference terms are sufficiently large and of the appropriate sign. 

Similar statements apply to the more general quasi-probability,
\beq
q(s_1, s_2, \cdots s_n) = {\rm Re} {\rm Tr} \left( P_{s_n} (t_n) \cdots P_{s_2} (t_2) P_{s_1} (t_1) \rho \right),
\label{quasin}
\eeq
and in particular, when negative, non-zero interference must be present. 
The condition
\beq
q(s_1, s_2, \cdots s_n ) \ge 0,
\eeq
is the quantum-mechanical counterpart of the condition Eq.(\ref{ntimes}).

The above quantities are simply related to the ``coherence witness" measure,
\beq
W(s_2) \equiv p_2 (s_2) - \sum_{s_1} p_{12} (s_1, s_2),
\eeq
which is clearly a measure of the degree to which the NSIT condition Eq.(\ref{NSIT12}) is violated.
It is readily seen from Eq.(\ref{A1}) to be equal to the sum of off-diagonal terms of the decoherence functional and we have that,
\beq
p_{12}(s_1,s_2) = q(s_1, s_2) - \frac{1}{2} W (s_2).
\label{A5}
\eeq
Since $p_{12}(s_1,s_2) \ge 0 $, a consequence of this is that for $q(s_1, s_2) \ge 0$ (i.e for the two-time LG inequalities to hold) we need either $W(s_2) \ge  0 $, 
or $W(s_2) < 0 $ with 
\beq
| W(s_2)| \le 2 p_{12}(s_1,s_2) 
\label{A6}
\eeq

This allows us to compare the two-time LG inequality violation with the condition Eq.(\ref{ineqp12}), since this condition may be written,
\beq
q(s_1, s_2) - \frac{1}{2} W (s_2) \le q(s_1, s_2) + q(-s_1,s_2).
\eeq
Cancelling an identical term from both sides and using Eq.(\ref{A5}) again, we have
\beq
p_{12}(s_1, s_2)  + W(s_2) \ge 0,
\eeq
for all $s_1$. This means that either, $W(s_2) \ge  0 $, 
or $W(s_2) < 0 $ with $| W(s_2)| \le  p_{12}(s_1,s_2) $, an upper bound half that in Eq.(\ref{A6}).
This means that firstly, the condition Eq.(\ref{ineqp12}) is violated by sufficiently large interference terms and secondly,  that it can be violated with the two-time LG inequalities still satisfied. Hence Eq.(\ref{ineqp12}) is a stronger condition than the two-time LG inequalities.

\bibliography{apssamp}

\begin{thebibliography}{10}








\bibitem{LG1} A.J.Leggett and A.Garg,
Phys. Rev. Lett. 54, 857 (1985).

\bibitem{L1} A. J. Leggett,
Foundations of Physics, 18, 939 (1988); 
Rep. Prog. Phys. 71, 022001 (2008).



\bibitem{HalQ} J.J.Halliwell, Phys. Rev. A 93, 022123 (2016).

\bibitem{HalLG4} J.J.Halliwell, Phys. Rev. A 96, 012121 (2017). A concise and updated version of this work is the e-print arXiv:1811.10408.
 


\bibitem{Fine} A.Fine, J.Math.Phys. 23, 1306 (1982); A.Fine, Phys.Rev.Lett. 48, 291 (1982).

\bibitem{HalFine} J.J.Halliwell, Phys.Lett. A378, 2945 (2014).



\bibitem{Laf} S.Majidy, H.Katiyar and R.Laflamme (private communication).

\bibitem{ELN}  C. Emary, N. Lambert and F. Nori, Rep. Prog. Phys. 77, 016001 (2014)


\bibitem{MaTi} O.J.E Maroney and C.G Timpson, arXiv:1412.6139 (2014).




\bibitem{KoBr} J. Kofler and C. Brukner,
Phys. Rev. A 87, 052115 (2013).


\bibitem{Cle} L.Clemente and J.Kofler, Phys. Rev. A 91, 062103 (2015); Phys. Rev. Lett. 116, 150401 (2016).

\bibitem{Ma1} O.Maroney (unpublished).

\bibitem{Wit} 
C.-M. Li, N. Lambert, Y.-N. Chen, G.-Y. Chen, and F. Nori,
Sci. Rep. 2, 885 (2012);
K. Wang, G. C. Knee, X.  Zhan, Z. Bian, J.  Li and P. Xue,
Phys. Rev. A 95, 032122 (2017).

\bibitem{ScEm} G. Schild and C.Emary, Phys. Rev. A 92, 032101 (2015).


\bibitem{WLG} D. Saha, S. Mal, P. K. Panigrahi and D. Home, Phys. Rev. A, 91, 032117 (2015);
S. Kumari and A. K. Pan, 
Phys. Rev. A 96, 042107 (2017); S. Kumari and A. K. Pan, 
EPL, 118, 50002 (2017).


\bibitem{Mon} A.Montina, Phys. Rev. Lett. 108, 160501 (2012).

\bibitem{Ye1} J.M.Yearsley, arXiv:1310.2149 (2013).

\bibitem{Guh}
O.G\"uhne, M.Kleinmann, A.Cabello, J-A. Larsson, G.Kirchmair, F.Z\"ahringer, R. Gerritsma and C. F. Roos,
Phys. Rev. A 81, 022121 (2010).






\bibitem{Knee} G. C. Knee, S. Simmons, E. M. Gauger, J. J. Morton, H. Riemann,
N. V. Abrosimov, P.Becker, H.-J. Pohl, K. M. Itoh, M. L. Thewalt, G. A. D. Briggs, and S. C. Benjamin, Nat. Commun. 3, 606 (2012).


\bibitem{Rob} 
C.Robens, W.Alt, D.Meschede, C.Emary and A.Alberti,
Phys. Rev. X 5, 011003 (2015).


\bibitem{KBLL} 
H. Katiyar, A. Brodutch, D. Lu, and R. Laflamme,
New J. Phys. 19, 023033 (2017).

\bibitem{EmINRM} 
K. Wang, C. Emary, X. Zhan, Z. Bian, J. Li and P. Xue,  
Opt. Exp. 25, 31462 (2017).


\bibitem{HalLG2} J.J.Halliwell, Phys. Rev. A 94, 052131 (2016).

\bibitem{HalLG3} J.J.Halliwell, Phys. Rev. A 94, 052114 (2016).

\bibitem{Ema} C.Emary, Phys. Rev. A 96, 042102 (2017).


\bibitem{Hue} S. F. Huelga, T. W. Marshall, and E. Santos, 
Phys. Rev. A 52, R2497 (1995);
S. F. Huelga, T.W. Marshall, and E. Santos, 
Phys. Rev. A 54, 1798 (1996);
S. F. Huelga, T. W. Marshall, and E. Santos, 
Europhys, Lett. 38, 249 (1997).

\bibitem{FA} S. Foster and A. Elby, Found. Phys. 21, 773 (1991).

\bibitem{AF} A. Elby and S. Foster, Phys. Lett. A 166, 17 (1992).

\bibitem{Lap} R. Lapiedra, Europhys, Lett. 75, 202 (2006).

\bibitem{Zuk} M. Zukowski, arxiv:1009.1749 (2010).


\bibitem{deco1} M.M.Wilde and A.Mizel, Found.Phys 42, 256 (2012).

\bibitem{deco2}E. Huffman and A. Mizel, Phys. Rev. A 95, 032131 (2017).

\bibitem{Dicke} R. H. Dicke,
Am. J. Phys 49, 925 (1981).

\bibitem{3box} R. E. George, L. M. Robledo, O. J. E. Maroney, M. S. Blok,
H. Bernien, M. L. Markham, D. J. Twitchen, J. J. L. Morton,
G. A. D. Briggs, and R. Hanson, 
Proc. Natl. Acad. Sci. USA 110, 3777 (2013).

\bibitem{Knee2} G. C. Knee, K. Kakuyanagi, M.-C. Yeh, Y. Matsuzaki, H. Toida,
H. Yamaguchi, S. Saito, A. J. Leggett, and W. J. Munro, 
Nat. Commun. 7, 13253 (2016).


\bibitem{Bec} A. Bechtold, F. Li, K. M\"uller, T. Simmet, P.-L. Ardelt, J. J. Finley, and N. A. Sinitsyn, Phys. Rev. Lett. 117, 027402 (2016).



\bibitem{Knee3} 
G. C. Knee, M. Marcus, L. D. Smith and A. Datta, Phys. Rev. A 98, 052328 (2018).

\bibitem{Wang} K. Wang, G. C. Knee, X. Zhan, Z. Bian, J. Li, and
P. Xue, Phys. Rev. A 95, 032122 (2017).


\bibitem{PaMa} J. P. Paz and G. Mahler,
Phys. Rev. Lett. 71, 3235 (1993).



\bibitem{SOS}
A. M. Souza, I. S. Oliveira and R. S. Sarthour,
New J. Phys. 13, 053023 (2011).

\bibitem{KGBB}
G. C. Knee, E. M. Gauger, G. A. D. Briggs and S. C. Benjamin,
New J. Phys. 14, 058001 (2012).

\bibitem{LGn1} D. Avis, P. Hayden, and M. M. Wilde, 
Phys. Rev. A 82, 030102 (2010).

\bibitem{LGn2} M. Barbieri, 
Phys. Rev. A 80, 034102 (2009).


\bibitem{HOM} A.Berdnorz and W.Belzig, Phys. Rev. B 81, 125112 (2010);
Phys. Rev. B 83, 125304 (2011). 






























\end{thebibliography}

\end{document}